\documentstyle[aps,prl,twocolumn,epsf]{revtex}
\begin{document}
\draft
\title{Constraining neutrino decays with CMBR data}
\author{Steen Hannestad}
\address{Theoretical Astrophysics Center, Institute of Physics and Astronomy,
University of Aarhus, 
DK-8000 \AA rhus C, Denmark}
\date{\today}
\maketitle

\begin{abstract}
The decay of massive neutrinos to final states containing only invisible
particles is poorly constrained experimentally. 
In this letter we describe the constraints that
can be put on neutrino mass and lifetime using CMBR measurements. 
We find that very tight lifetime limits 
on neutrinos in the mass range 10 eV - 100 keV can be
derived using CMBR data from upcoming satellite measurements. 
\end{abstract}


If the mass of any given neutrino species is larger than roughly 100 eV
it is necessarily unstable with a lifetime shorter than the Hubble 
expansion timescale \cite{cowsik}. 
A massive neutrino could in principle have many
different allowed decay modes, but if the final state contains 
electromagnetically interacting particles the decay is
ruled out by observations \cite{radobs}, unless the lifetime is very
long as in the scenario proposed by Sciama \cite{sciama}. 
However, if the final state contains only
``invisible'' particles (massless neutrino species, majorons, etc.)
it is very difficult to constrain the decay by looking for the decay
products, and in this case it becomes important to use indirect arguments.
In this letter we shall look at this latter type of decay, where the
decay products cannot be directly observed.

By using arguments pertaining to big bang nucleosynthesis one can rule
out a large chunk of the mass-lifetime parameter space \cite{nucleo}. 
In essence, the
argument is that if the mass is too large and the lifetime too long,
there is too much energy density present in the universe during 
nucleosynthesis which in turn leads to an overproduction of helium 
\cite{nucleo}.
If the heavy neutrino is meta stable during nucleosynthesis, one can rule
out a neutrino mass larger than a fraction of an MeV, but if the neutrino
is unstable on the timescale of nucleosynthesis ($\tau \lesssim 10^3$s),
the argument becomes more complicated, and certain lifetimes are allowed
even for heavy neutrinos.
Altogether, nucleosynthesis can be used to constrain lifetimes for
masses larger than roughly 0.1 MeV \cite{nucleo,hannestad}.

In the present letter we focus on another method that can be used
for the same purpose. The cosmic microwave background radiation
(CMBR) is created at
the last scattering surface for photons. 
The CMBR is anisotropic at the $10^{-5}$ level, a fact long predicted
but only detected in 1992 by the COBE satellite \cite{COBE}. 
The physical conditions
prevailing at the epoch of last scattering are imprinted in these
anisotropies, meaning that one can in principle determine the physical
content of the universe at photon decoupling by measuring the anisotropy
spectrum \cite{tegmark}.
Photons decouple from the cosmic plasma at a temperature
of approximately 1 eV and if conditions are changed (for example by
a heavy, decaying neutrino) prior to this, it is possibly noticeable
in the CMBR anisotropies. 
Therefore it should be possible to use such arguments
to constrain decays happening as late as at a temperature in the eV 
range.
The influence of decaying neutrinos on structure formation have been
considered many times in the literature \cite{decstruc,silk} and it is also
well known that decaying neutrinos may influence the 
CMBR power spectrum \cite{silk}. 
However, such calculations have not focussed on the limits that may be
put on neutrino masses and lifetimes. Previously, the data have not
been sufficient, but within the next few years 
we can expect to have extremely fine data for the power spectrum
of microwave anisotropies, measured by the upcoming satellite missions
MAP and PLANCK \cite{MAP+PLANCK}. This is the primary motivation
for discussing the neutrino mass and lifetime limits that can possibly be
obtained from CMBR data.

We shall here assume that the decaying neutrino is the $\tau$ neutrino
and that the two lighter neutrinos are massless (that is, they have
masses much smaller than the photon decoupling temperature). The same type of 
argument could also be used for an unstable muon neutrino, but if
the muon neutrino is massive and decays on this timescale, it would
also imply a heavy and unstable tau neutrino unless there is some
kind of inverted mass hierarchy. Having
a scenario with two decaying neutrino species complicates the matter
unnecessarily although such a scenario could have interesting 
consequences \cite{lasing}.
The decay we study can therefore be written generically as
\begin{equation}
\nu_\tau \to D,
\end{equation}
where $D$ indicates the final state ($\nu \bar\nu \nu$,
$\nu \phi$, etc.) configuration.
We only look at decays taking place after Big Bang nucleosynthesis;
that is, we assume $m_{\nu_\tau} \wedge T(\tau) \lesssim 0.1$ MeV.

If the decay final state contains only massless particles and we ignore
possible inverse decays as well as relativistic corrections the
equations describing the decay simplify a lot. We follow for instance
Dodelson, Gyuk and Turner \cite{DGT94} in taking this approach. 
In that case, the
cosmological evolution equations have the form
\begin{eqnarray}
\rho_{\nu_\tau} & = &\rho_0 \frac{\sqrt{(3.151T_0)^2+m^2}}{3.151T_0}
\exp(-t/\tau), \nonumber \\
\dot{\rho}_D & = & -4 H \rho_D + \rho_{\nu_\tau}/\tau, \nonumber \\
\rho_0 & = &\frac{7}{120} \pi^2 T_0^4, \label{eq:evol} \\
\rho_\gamma & = & \left(\frac{11}{4}\right)^{4/3}\frac{8}{7} \rho_0, 
\nonumber\\
\rho & = & \rho_{\nu_\tau} + 2 \rho_0 + \rho_D + \rho_\gamma,\nonumber \\
\frac{\dot{T_0}}{T_0} & = & - \sqrt{\frac{8\pi G \rho}{3}},\nonumber
\end{eqnarray}
where $T_0$ and $\rho_0$ indicate the temperature and energy density
of a standard massless neutrino species. The subscript $D$ indicates
the decay products and $\tau$ is the lifetime of $\nu_\tau$.
We assume that the decay is finished well before the
last scattering (in practice we assume that neither the heavy neutrino
mass nor the temperature corresponding to $\tau$ are smaller than 10 eV).
This means that the decay only perturbs the CMBR through the change
in relativistic energy density at last scattering. 
By using these assumptions the decay simplifies further because the
final relativistic energy density depends on $m_{\nu_\tau}$ 
and $\tau$ only through
the combination $m_{\nu_\tau}^2 \tau$. One can define a decay ``relativity''
parameter, $\alpha$, as \cite{lasing}
\begin{equation}
\alpha \equiv 3.50 \, \left(\frac{m_{\nu_\tau}}{1{\rm keV}}\right)^2 
\left(\frac{\tau}{1{\rm y}}
\right),
\end{equation}
the decay being relativistic roughly for $\alpha \lesssim 1$ and 
non-relativistic for $\alpha \gtrsim 1$.
We follow the
standard practice in expressing the relativistic energy density in
equivalent number of massless neutrino species \cite{kolbturner}
\begin{equation}
N_\nu \equiv (2 \rho_0 + \rho_D)/\rho_0.
\end{equation}
Within our approximation, $N_\nu \to 3$ for relativistic decays, since
in that case all that really happens is a reshuffling of relativistic
energy density between different species. If one includes
inverse decays, relativistic decays proceed in equilibrium so that
the decay does not finish before the heavy neutrino goes non-relativistic.
This means that even for strongly relativistic decays there will be
some contribution to the energy density of the decay products from the
rest mass term, so that $N_\nu > 3$ in this case. The decay formalism
is much more complicated, however, since one then has to use a specific
final state (depending on the number of final state particles and their
quantum statistics) and solve the full Boltzmann equation \cite{lasing}. 
For this reason we just use the simplistic approach
because the main point of the letter is not to derive precision limits
but rather to point out the strength of this type of argument.

For very non-relativistic decays one can also arrive at an analytic
approximation for $N_\nu$. Assuming that the cosmic energy density
just prior to decay is completely dominated by the heavy neutrino and
that the decay proceeds instantaneously at $t=\tau$, one gets
$N_\nu \simeq 0.52 \, \alpha^{2/3}$.
Altogether we can approximate $N_\nu$ for any value of $\alpha$ with
the function
\begin{equation}
N_{\nu,{\rm app}} \simeq 3 + 0.52 \, \alpha^{2/3},
\label{eq:approx}
\end{equation}
which has the correct asymptotic behaviour. In Fig.~1 we have plotted
both $N_\nu$ from the numerical solution of Eq.~(\ref{eq:evol}) and
from Eq.~(\ref{eq:approx}). In the remainder of the
letter we shall use our analytic approximation, Eq.~(\ref{eq:approx}),
to simplify calculations.
% --------------------------------------
% Figure 1
% --------------------------------------
\begin{figure}[t]
\begin{center}
\epsfysize=7truecm\epsfbox{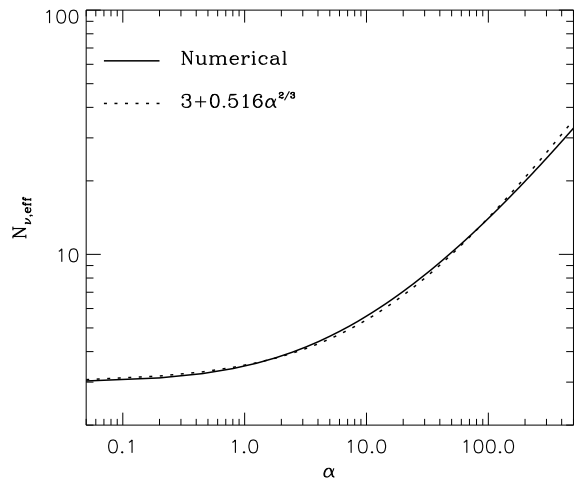}
\vspace{0truecm}
\end{center}
\baselineskip 17pt
\caption{The equivalent number of neutrino species produced by decay.
Both the solution obtained by use of Eq.~(\ref{eq:evol}) and our
analytic approximation, Eq.~(\ref{eq:approx}), are shown.}

\label{fig1}
\end{figure}
Note here, that we have only included the standard three neutrino species
and the decay products in this calculation. In principle there could
be energy density contributions from other more exotic particles,
which would add some constant value to $N_\nu$. 
We shall not deal further
with this possibility here, only note that if this is the case, the
allowed value of $\alpha$ corresponding to a given $N_\nu$ is shifted to 
lower values.

The microwave background anisotropies are usually calculated in terms
of $C_l$ coefficients that relate to the fluctuation spectrum as
$C_l \equiv \langle|a_{lm}|^2\rangle$,
where $a_{lm}$ can be expressed in terms of the temperature fluctuations
as
$T(\theta,\phi) =\sum_{lm}a_{lm}Y_{lm}(\theta,\phi)$.
At present, CMBR data are quite insufficient to put any constraints on
the equivalent number of neutrino species. The main reason is that the
only real precision data at present are the COBE data \cite{COBE}. 
This data only
gives the power spectrum at $l \lesssim 20$, where the spectrum is
flat and featureless.
However, as previously mentioned this can be expected
to change dramatically within the next few years, where two different 
satellite missions are expected to measure the CMBR power spectrum to
high precision out to values of $l$ beyond 1000 \cite{MAP+PLANCK}. 
This will allow
for determination of almost all of the relevant cosmological 
parameters to great precision, including the number of light
neutrino species \cite{JKKS95}.
% --------------------------------------
% Figure 2
% --------------------------------------
\begin{figure}[h]
\begin{center}
\epsfysize=7truecm\epsfbox{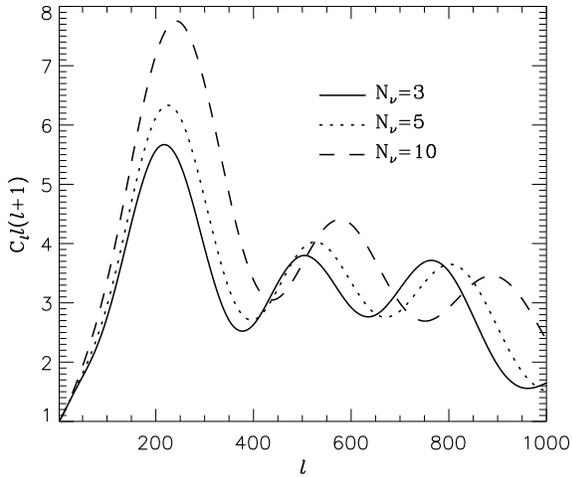}
\vspace{0truecm}
\end{center}
\baselineskip 17pt
\caption{CMBR power spectra for models with different values of $\alpha$.
The spectra have been normalised to the quadrupole coefficient, $6C_2$.
The underlying model is a standard CDM model with $\Omega=1$,
$\Omega_b=0.05$ and $h=0.5$.}

\label{fig2}
\end{figure}

Several different groups have performed calculations of the precision
to which one can measure different parameters, depending on the
quality of the data \cite{JKKS95,tegmark2}. 
It turns out that the CMBR power spectrum is very sensitive to $N_\nu$
because varying the relativistic energy density also means shifting
matter-radiation equality.
In the matter dominated epoch photons see a constant gravitational
potential (in time) after they are emitted from the last scattering surface,
at least in the linear approximation. If the universe is not 
completely matter dominated at last
scattering the potential is not constant and this in turn increases
the CMBR anisotropy, an effect known as the early ISW effect
\cite{tegmark}. 
If the amount of relativistic energy density is increased, last scattering
will occur closer to the radiation dominated epoch, meaning that 
especially the first Doppler peak is increased in height.
In Fig.~2 we show the CMBR power spectrum for different values of
$\alpha$. Clearly, the first Doppler peak is very sensitive to changes
in $\alpha$, exactly as expected.
This high sensitivity to $N_\nu$ means that $N_\nu$ can be measured
fairly precisely by CMBR experiments.
As shown in Ref.~\cite{JKKS95}, experiments like MAP and PLANCK should
realistically be able to measure $\Delta N_\nu$ to within 0.3
without any prior knowledge of other cosmological parameters,
and to
much greater precision assuming some prior knowledge of other 
parameters.
Note here that by using nucleosynthesis arguments one
can also constrain $N_\nu$ to be below roughly 4, but this argument
only holds for temperatures above the range we are interested in.
In our case we assume that the decay takes place after nucleosynthesis
so that during nucleosynthesis we have $N_\nu=3$.

Now, if we assume a given mass for $\nu_\tau$, we can relate $\Delta N_\nu$
to an allowed lifetime interval $\Delta \tau$ for any value of $N_\nu$. 
Doing this gives
\begin{equation}
\tau ({\rm y})^{\pm} = \left(\frac{N_\nu \pm \Delta N_\nu-3}
{1.19}\right)^{3/2} 
m^{-2}({\rm keV}),
\end{equation}
where $\tau^+$ indicates the maximum and $\tau^-$ the minimum allowed 
lifetime.
% --------------------------------------
% Figure 3
% --------------------------------------
\begin{figure}[h]
\begin{center}
\epsfysize=7truecm\epsfbox{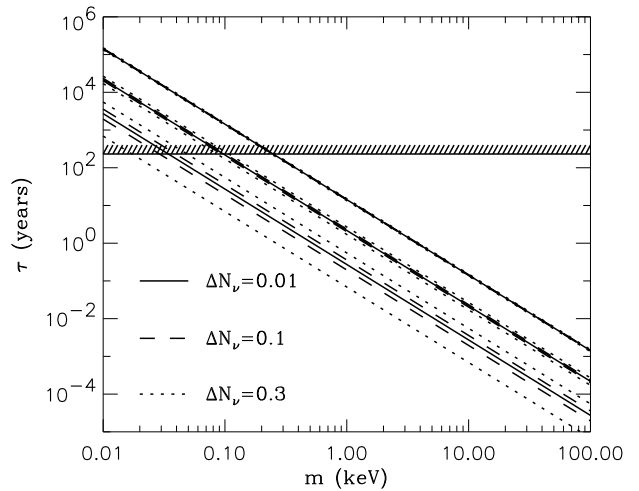}
\vspace{0truecm}
\end{center}
\baselineskip 17pt
\caption{Allowed lifetime intervals for neutrino decay, plotted for 
different values of measured $N_\nu$. The three intervals correspond to
$N_\nu =$ 3.5, 5 and 10; the interval with the longer lifetime
corresponding to the higher $N_\nu$.}

\label{fig3}
\end{figure}
In Fig.~3 we have shown the allowed lifetime intervals for a decaying
tau neutrino as a function of neutrino mass for different measured
values of $N_\nu$ and assuming different precision. Especially for
non-relativistic decays one can constrain the lifetime to lie within
a very narrow band. Note that measuring $N_\nu \gg 3$ is of course not
a confirmation of neutrino decay, although that would be a possible
explanation for such an effect. As such, the CMBR arguments (like
nucleosynthesis arguments) can only serve to rule out certain regions
of parameter space, not provide an unambiguous detection.

In conclusion, we have shown that the CMBR is a very sensitive 
tool for constraining
neutrino decays, and with new CMBR data it should become possible to
narrow down the allowed lifetime intervals for neutrino masses in
the range of 10 eV to 100 keV.
The constraints coming from this type of argument are applicable to
any type of neutrino decays to massless, non-interacting final states.
What is constrained is really the amount of relativistic energy density
at last scattering, and in that regard this type of argument is very
similar to nucleosynthesis constraints. 
What is different with the CMBR argument is that it pertains to a
completely different mass-lifetime region. Here, we constrain masses
in the region below 0.1 MeV, so that by putting together CMBR and 
nucleosynthesis arguments essentially {\it all} masses down to $O(10$eV)
are covered, since nucleosynthesis already deals with more or less
all masses up to the maximum experimentally allowed mass of 
24 MeV \cite{mnutau}.
Thus, although CMBR data cannot at present constrain neutrino decays
to invisible final states it is to be expected that in the very 
near future this will change. One should then be able to constrain such
neutrino decays very strongly with new data. 
\vspace*{0.5cm}\\
This work was supported by the Theoretical Astrophysics Center under
the Danish National Research Foundation.

\end{document}